\documentclass[11pt]{article}

\usepackage[dvips]{graphics}
\usepackage{epsfig}
\usepackage[caption=false,font=footnotesize]{subfig}

\usepackage{amsmath,amssymb}
\usepackage{bm,color}

\usepackage[round]{natbib}

\newcommand{\tetn}{{\mbox{\boldmath $\theta$}}}

\title{Bayesian Estimation of the Kumaraswamy Inverse Weibull Distribution}

\author{Felipe R. S. de Gusm\~{a}o, Vera L. D. Tomazella and Ricardo S. Ehlers}

\begin{document}\maketitle

\begin{abstract}
  
The Kumaraswamy Inverse Weibull distribution has the ability to model
failure rates that have unimodal shapes and are quite common in
reliability and biological studies. The three-parameter Kumaraswamy
Inverse Weibull distribution with decreasing and unimodal failure rate
is introduced. We provide a comprehensive treatment of the
mathematical properties of the Kumaraswany Inverse Weibull
distribution and derive expressions for its moment generating function
and the $r$-th generalized moment. Some properties of the model with
some graphs of density and hazard function are discussed. We also discuss
a Bayesian approach for this distribution and an application was made
for a real data set.
\vspace{0.2cm} 


\end{abstract}

\section{Introduction}

In this paper we propose a new probability distribution to handle the
problem of survival data. Motivated by research developed in recent
years, we introduce the Kumaraswany Inverse Weibull distribution that
includes several well known distributions used in survival analysis. 

Recently, many authors have proposed new classes of distributions, which
are modifications of distribution functions which provide hazard
ratios contemplating various shapes. We can cite for example the Weibull
exponential \citep{Mudholkar1995}, which has also the hazard rate function
with a unimodal form,  (see also \citet{Xie1996}). 
\cite{Carrasco2008} proposed  
a four-parameter distribution denoted generalized modified Weibull
(GMW) distribution, \cite{Gusmao2011} introduced and studied the
tri-parametric inverse Weibull generalized  distribution that
possesses failure rate with unimodal, increasing and decreasing
forms. \cite{Pescim2010} proposed a distribution with four parameters,
called beta generalized half normal distribution. 

Underexplored in the literature and rarely used by statisticians, the
Kumaraswamy distribution \citep{Kumaraswamy1980} has a domain in the
real interval $(0,1)$. This property turns the Kumaraswamy
distribution a natural candidate to combine with other distributions to
produce a more general one. Its cumulative distribution function (cdf) is
given by, 
\begin{equation}\label{eq_kum-cdf}
F\left(x; a, b \right)= 1- \left[ 1- x^a \right]^b,
\end{equation}
and its probability density function (pdf) is given by,
\begin{equation*}
f\left(x; a, b \right)=a b  x^{a-1} \left[ 1-x^a\right]^{b-1}, ~0<x<1,
\end{equation*}
where $a>0$ and $b>0$. This density can be unimodal, increasing,
decreasing or constant. 

Recently, \cite{Cordeiro2011} proposed to use the Kumaraswamy to generalize
other distributions. Considering that a random variable $X$ has
distribution $G$, they suggest to apply the Kumarasawamy distribution
to $G(x)$. Note that, since $0<G(x)<1$ for any distribution function
$G$, then evaluating equation (\ref{eq_kum-cdf}) at $G(x)$ we obtain, 
\begin{equation}\label{eq_KG}
F_G(x; a, b)= 1 - [1 - G(x)^a]^b.
\end{equation}
where $F_G$ is the cdf of the generalized Kumaraswamy-$G$ distribution.
Based on these ideas, we consider the Inverse Weibull Distribution as
a candidate for $G$, using Equation (\ref{eq_KG}). Then, performing some
adjustments and mathematical manipulations, we obtain the Kumaraswamy
inverse Weibull (Kum-IW) distribution. 

The rest of the paper is organized as
follows. In Section \ref{sec_Kum-IW}, we develop the Kum-IW
distribution. Section \ref{sec:properties} is devoted to describe
basic properties of the distribution.
Inference procedures via maximum
likelihood and Bayesian approaches are presented in 
Section \ref{sec:estimation}. Section \ref{Application}
is devoted to analyze a real data set and in Section \ref{sec_final}
we present some conclusions of this work. 

\section{Kumaraswamy inverse Weibull distribution}\label{sec_Kum-IW}

Let $T$ a random variable with inverse Weibull distribution. Then its cdf can be written as,
\begin{equation}\label{eq_IW}
G(t) = \exp\left[- \left(\frac{\alpha}{t} \right)^{\beta}\right], \quad t > 0,
\end{equation}
where $\alpha > 0$, $\beta > 0$, and its pdf is given by,
\begin{eqnarray*}
g(t) = \beta \alpha^{\beta}t^{-\left(\beta+1\right)}\exp\left[-\left(\frac{\alpha}{t}\right)^{\beta}\right].
\end{eqnarray*}

\noindent Inserting the $G$ function of Equation (\ref{eq_IW}) in
Equation (\ref{eq_KG}) it follows that,
\begin{equation}\label{eq_Kum-IW-1}
F_G(t; a, b, \alpha, \beta) = 1-\left\{ 1-\exp \left[-a\left( \frac{\alpha}{t} \right)^\beta \right] \right\}^b.
\end{equation}

\noindent We note that the parameters $a$ and $\alpha$ in (\ref{eq_Kum-IW-1})
are not identifiable and we adopt the reparameterization $c=\alpha
a^{1/\beta}$ so that the Kum-IW cdf is rewritten as,
\begin{equation}\label{eq_Kum-IW}
F_G(t; b, c, \beta) = 1- \left\{ 1-\exp \left[ -\left( \frac{c}{t} \right)^\beta \right] \right\}^b,
\end{equation}
where $b > 0$ and $c>0$ are the shape and scale paremeters
respectively. 
Accordingly, the Kum-IW pdf is now given by,
\begin{equation}\label{eq_f-Kum-IW}
f_G(t; b, c, \beta) = \beta b c^{\beta} t^{-\left(\beta+1\right)} \exp \left[ -\left( \frac{c}{t} \right)^\beta \right]
\left\{ 1-\exp \left[ -\left( \frac{c}{t} \right)^\beta \right] \right\}^{b-1}.
\end{equation}
It can be easily seen that when $b=1$ we obtain the pdf of the Inverse Weibull (IW) distrbution given by,
\begin{eqnarray*}
g(t) = \beta \alpha^{\beta}t^{-\left(\beta+1\right)}\exp\left[-\left(\frac{\alpha}{t}\right)^{\beta}\right].
\end{eqnarray*}

Finally, the corresponding survival and hazard functions are respectively given by,
\begin{equation*}
S_G(t; b, c, \beta) = \left\{ 1-\exp \left[ -\left( \frac{c}{t} \right)^\beta \right] \right\}^b
\quad\mbox{and}\quad
h_G(t; b, c, \beta) = 
\frac{\beta b c^{\beta} t^{-\left(\beta+1\right)} \exp \left[ -\left( \frac{c}{t} \right)^\beta \right]}
{1-\exp \left[ -\left( \frac{c}{t} \right)^\beta \right]}.
\end{equation*}
while the quantile function, $Q(u)$, of the Kum-IW distribution is given by,
\begin{equation*}
Q(u) = F^{-1}(u; b, c, \beta) = c(-\log(1-(1-u)^{\frac{1}{b}}))^{-{\frac{1}{\beta}}}
\end{equation*}

\subsection{Some special classes of the Kum-IW}\label{sec_GBRM}

The following well known and new distributions are special sub-classes of the Kum-IW 
distribution.

\begin{itemize}
	
\item Kumaraswamy inverse Rayleigh distribution (Kum-IR)

If $\beta = 2$, the Kum-IW distribution reduces to the Kumaraswamy inverse Rayleigh distribution (Kum-IR). 
Then, with $\beta = 2$ the density function of Kum-IW is expressed by:
\begin{equation*}%\label{eq_KiR}
F_G(t; b, c) = 1- \left\{ 1-\exp \left[ -\left( \frac{c}{t} \right)^2 \right] \right\}^b,
\end{equation*}
where $b > 0$ is the shape parameter, and $c > 0$ is the scale
parameter. Hence, the KiR distribution has two parameters, and its pdf
is given by 
\begin{equation*}%\label{eq_f-KiR}
f_G(t; b, c) = 2 b c^{2} t^{-3} \exp \left[ -\left( \frac{c}{t} \right)^2 \right]
\left\{ 1-\exp \left[ -\left( \frac{c}{t} \right)^2 \right] \right\}^{b-1}.
\end{equation*}

The corresponding survival and hazard functions are given respectively
by,
\begin{equation*}
S_G(t; b, c) = \left\{ 1-\exp \left[ -\left( \frac{c}{t} \right)^2 \right] \right\}^b
~~~~\textnormal{and}~~~~
h_G(t; b, c) = \frac{2 b c^{2} t^{-3} \exp \left[ -\left( \frac{c}{t} \right)^2 \right]}
{1-\exp \left[ -\left( \frac{c}{t} \right)^2 \right]}.
\end{equation*} 

\item Inverse Rayleigh distribution (IR)

If $\beta = 2$ and $b=1$, the Kum-IW distribution reduces to the inverse Rayleigh distribution (Kum-IR). 
Then, with $\beta = 2$ and $b=1$ the density function of Kum-IW is expressed by:
\begin{equation*}%\label{eq_iR}
G(t) = \exp\left[- \left(\frac{\alpha}{t} \right)^{2}\right], \quad t > 0,
\end{equation*}
where $\alpha > 0$, and its pdf is
\begin{eqnarray*}
g(t) = 2 \alpha^{2}t^{-3}\exp\left[-\left(\frac{\alpha}{t}\right)^{2}\right].
\end{eqnarray*}

\item Kumaraswamy inverse Exponential distribution (Kum-IE)
	
If $\beta = 1$, the Kum-IW distribution reduces to the Kumaraswamy inverse Exponential distribution (Kum-IE). 
Then, with $\beta = 1$ the density function of Kum-IW is expressed by:
\begin{equation*}%\label{eq_KIE}
F_G(t; b, c) = 1- \left\{ 1-\exp \left[ -\left( \frac{c}{t} \right) \right] \right\}^b,
\end{equation*}
where $b > 0$ is the shape parameter, and $c > 0$ is the scale
parameter. Hence, the KIE distribution has two parameters, and its pdf
is given by 
\begin{equation*}%\label{eq_f-KIE}
f_G(t; b, c) =  b c t^{-2} \exp \left[ -\left( \frac{c}{t} \right) \right]
\left\{ 1-\exp \left[ -\left( \frac{c}{t} \right) \right] \right\}^{b-1}.
\end{equation*}

The corresponding survival and hazard functions are respectively
\begin{equation*}
S_G(t; b, c) = \left\{ 1-\exp \left[ -\left( \frac{c}{t} \right) \right] \right\}^b
~~~~\textnormal{and}~~~~
h_G(t; b, c) = \frac{ b c t^{-2} \exp \left[ -\left( \frac{c}{t} \right) \right]}
{1-\exp \left[ -\left( \frac{c}{t} \right) \right]}.
\end{equation*}

\item Inverse Exponential distribution (IE)
	
If $\beta = 1$ and $b=1$, the Kum-IW distribution reduces to the Inverse Exponential distribution (IE). 
Then, with $\beta = 1$ and $b=1$ the density function of Kum-IW is expressed by:
	
\begin{equation*}%\label{eq_IE}
G(t) = \exp\left[- \left(\frac{\lambda}{t} \right)\right], \quad t > 0,
\end{equation*}
where $\lambda > 0$ and its pdf is
\begin{eqnarray*}
g(t) = \lambda t^{-2}\exp\left[-\left(\frac{\lambda}{t}\right)\right].
\end{eqnarray*}

\end{itemize}

\begin{figure}[h]\centering
\subfloat[\label{fig_1a}]{
  \includegraphics[width=6.5cm,height=6.5cm]{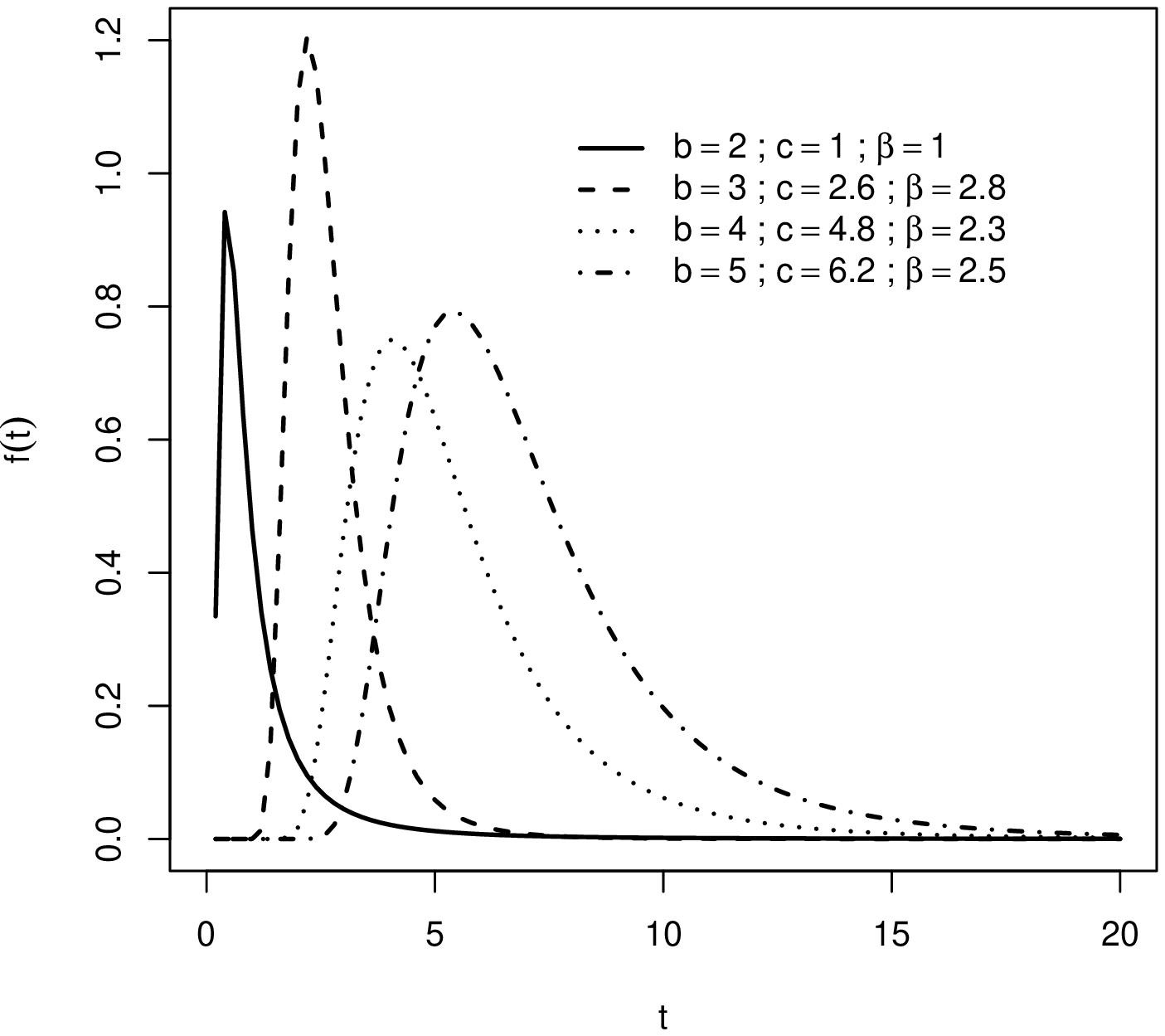}
}
\subfloat[\label{fig_1b}]{
  \includegraphics[width=6.5cm,height=6.5cm]{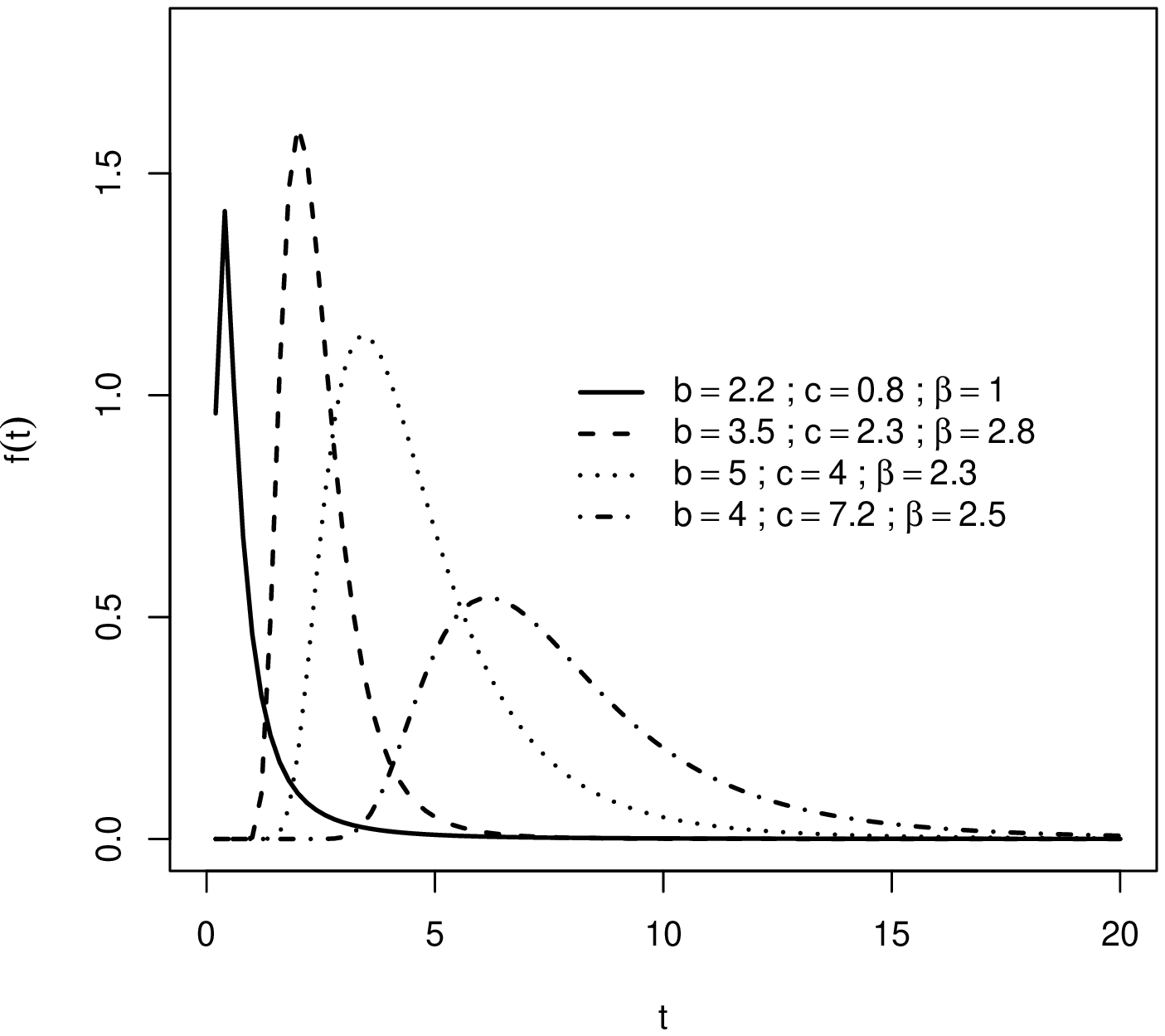}
}
\caption{\label{fig_Kum-IW-dens} Kumaraswamy inverse Weibull density functions.} 
\end{figure}

\begin{figure}[h]\centering
\subfloat[\label{fig_2a}]{
  \includegraphics[width=6.5cm,height=6.5cm]{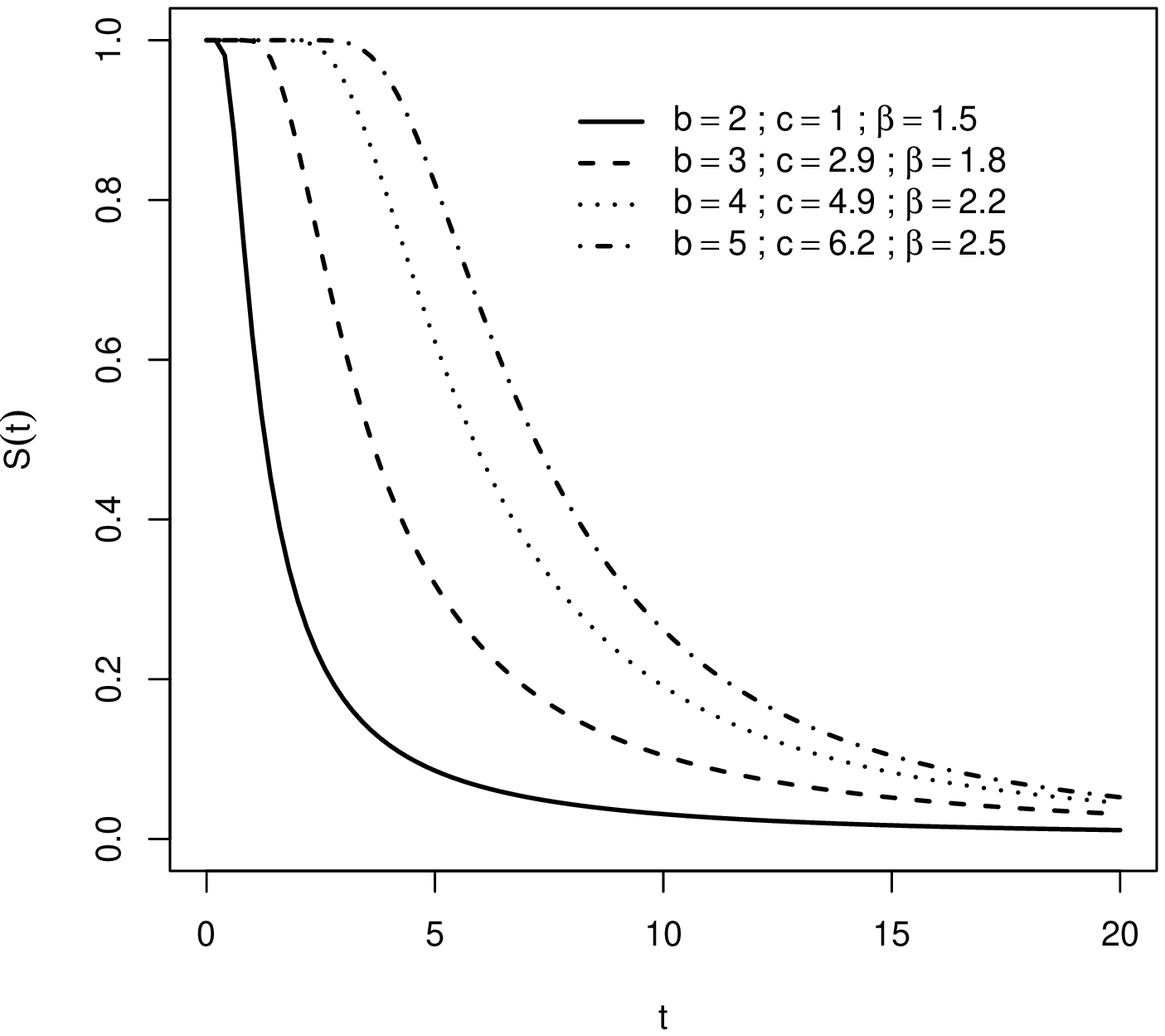}
}
\subfloat[\label{fig_2b}]{
  \includegraphics[width=6.5cm,height=6.5cm]{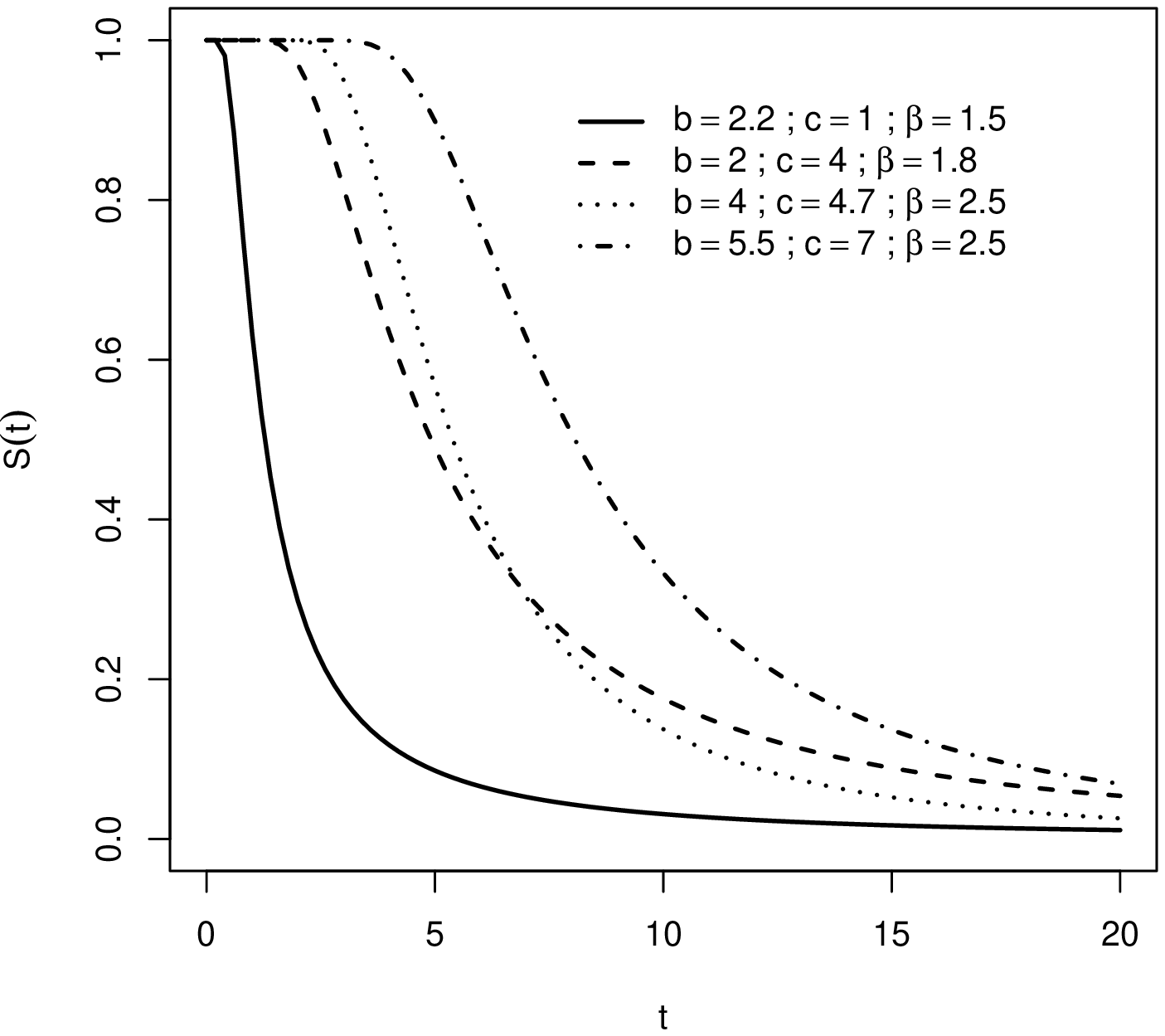}
}
\caption{\label{fig_Kum-IW-surv} Kumaraswamy inverse Weibull survival functions.}
\end{figure}

%\begin{center}
%Figure \ref{fig_Kum-IW-surv} about here.
%\end{center}

\begin{figure}[h]\centering
\subfloat[\label{fig_3a}]{
  \includegraphics[width=6.5cm,height=6.5cm]{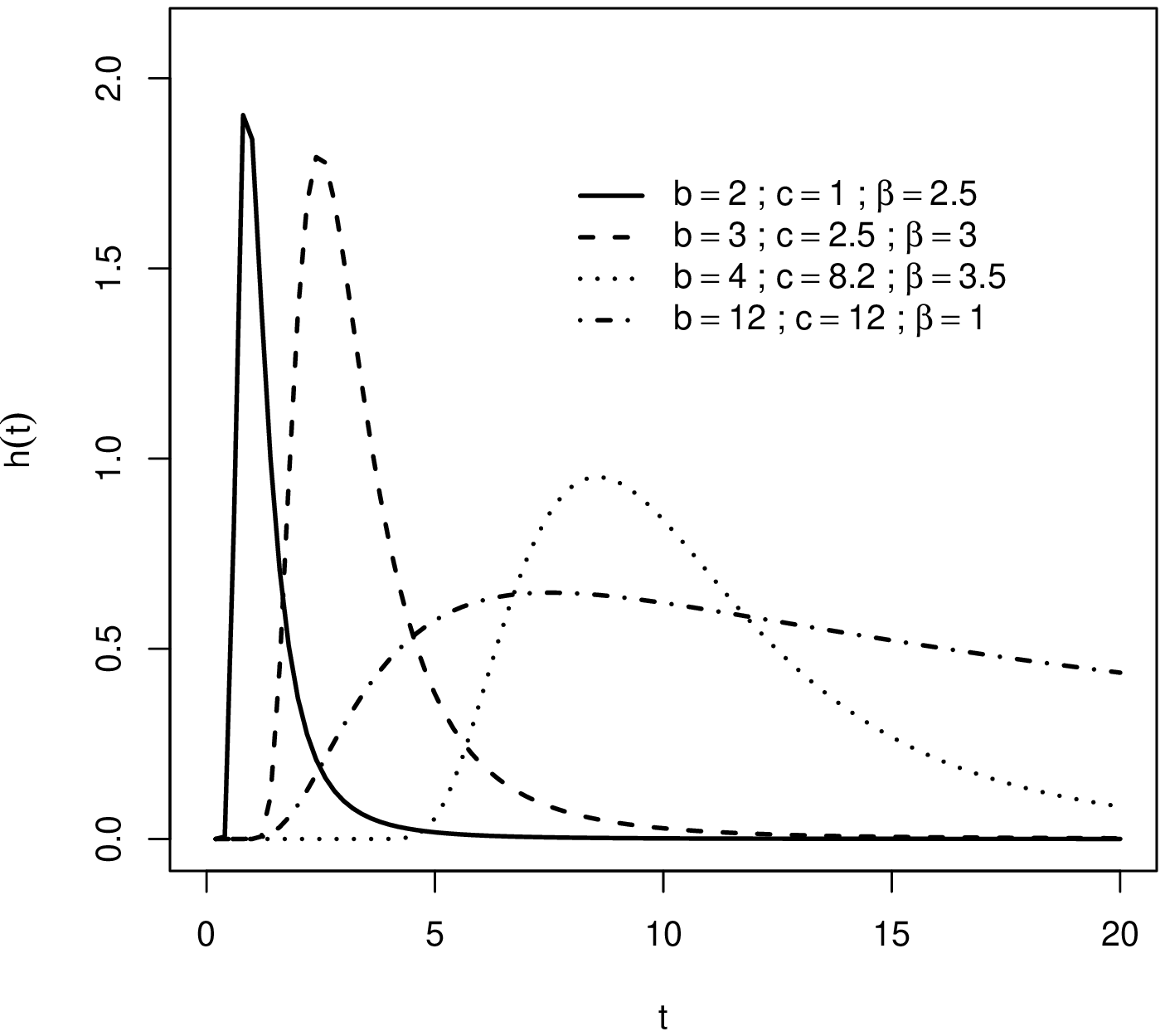}
}
\subfloat[\label{fig_3b}]{
  \includegraphics[width=6.5cm,height=6.5cm]{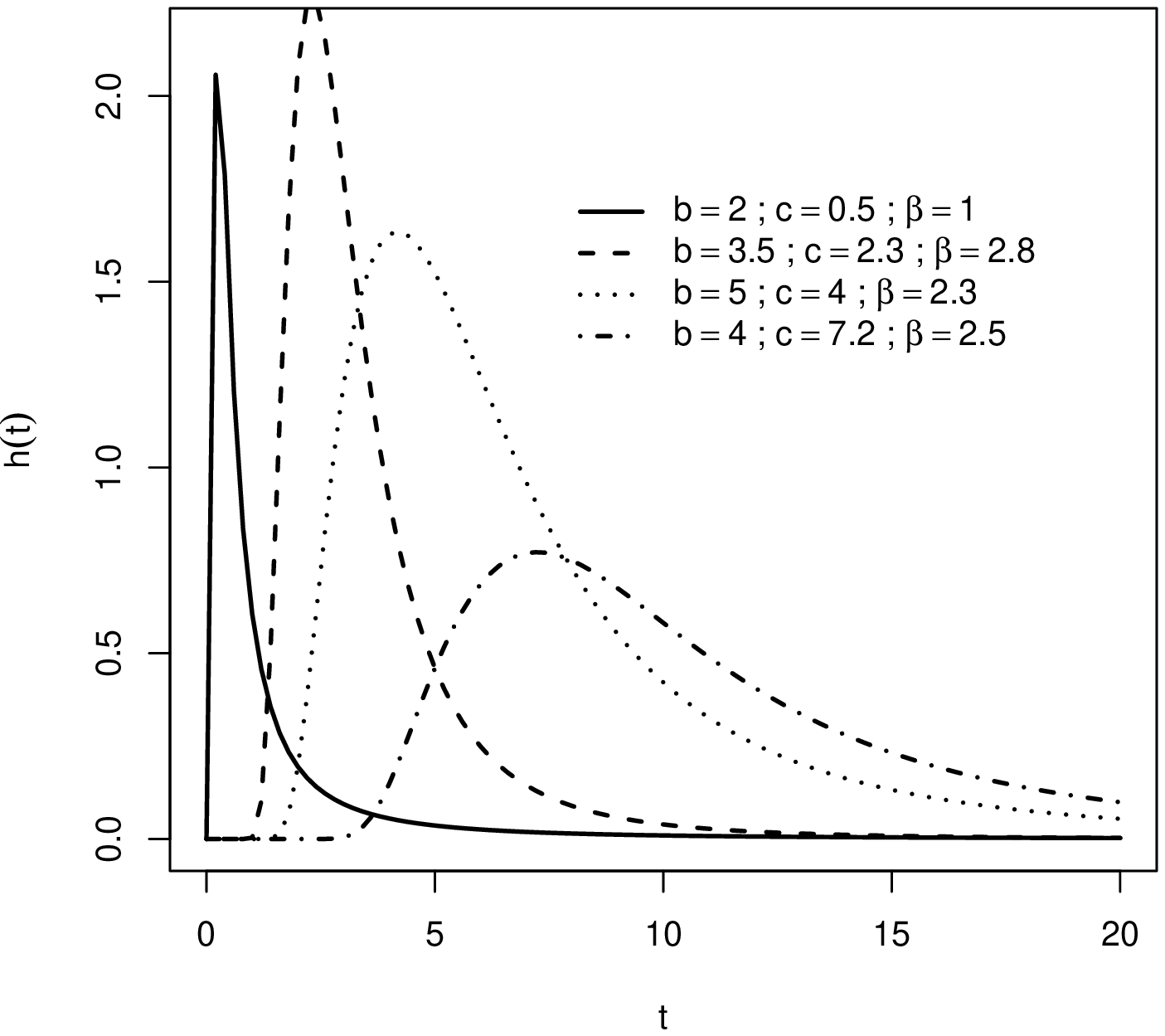}
}
\caption{\label{fig_Kum-IW-hazard} Kumaraswamy inverse Weibull hazard functions.}
\end{figure}

%\begin{center}
%Figure \ref{fig_Kum-IW-hazard} about here.
%\end{center}

\section{Basic Properties}\label{sec:properties}

In this section we describe in detail some properties like
expansions, moments, mean deviations,
Bonferroni and Lorenz curves, order statistics and entropies which
might be useful in any application of the distribution.

\subsection{Expansions for the distribution and density functions}
\label{sec_expan}

We now give simple expansions for the cdf of the Kumaraswamy Inverse
Weibull distribution. If $|x|< 1$ and $\psi > 0$ is 
a non-integer real number, we have
\begin {equation}\label{01}
\left(1-x\right)^{\psi}=\sum^{\infty}_{i=0}\left(-1\right)^{i} \left(\psi !\right) x^{i}.
\end{equation}

\noindent If $\psi$ is a positive integer, the series stops at $i=\psi$. Using
expansion in Equation (\ref{01}) it follows that,
\begin{equation}\label{1}
f\left(t; b, c, \beta \right)= \beta b c^{\beta}
t^{-\left(\beta+1\right)} \sum^{\infty}_{i=0}
\frac{\left(-1\right)^{i} \Gamma\left(b\right)}{\left(i !\right)
  \Gamma\left(b-i\right)} 
\exp \left[ -\left( \frac{c}{t} \right)^\beta \left(i+1\right) \right].
\end{equation}
and
\begin{equation*}%\label{1.1}
F\left(t; b, c, \beta \right)=1-\sum^{\infty}_{i=0}
\frac{\left(-1\right)^{i} \Gamma\left(b+1\right)}{\left(i !\right)
  \Gamma\left(b+1-i\right)} 
\exp \left[ -\left( i \frac{c}{t} \right)^\beta \right].
\end{equation*}

\noindent Because the integrals involved in the computation of moments, Bonferroni and Lorenz curves, 
reliability, Shannon and R\'enyi entropies and other inferential results do 
not have analytical solutions, these expansions are necessary.

\subsection{A general formula for the moments of the Kum-IW}\label{sec_mom}

We hardly need to emphasize the need and importance of the moments in any
statistical analyses, especially in applied work. Some of the most important features
and characteristics of a distribution can be studied using their moments (e.g. tendency,
dispersion, skewness and kurtosis). If the random variable $T$ follows
the Kum-IW distribution, its $k$-th moment about zero is given by, 
\begin{eqnarray*}
E\left(t^k\right) 
&=& 
\int^{\infty}_{0} t^{k} \beta b c^{\beta} t^{-\left(\beta+1\right)}
\exp \left[ -\left( \frac{c}{t} \right)^\beta \right] 
\left[ 1-\exp \left[- \left( \frac{c}{t} \right)^\beta \right] \right]^{b-1}dt\\
&=&
b c^k \sum^{\infty}_{r=0} \frac{\left(-1\right)^r \Gamma\left(b\right)}{\Gamma\left(b-r\right) r!} 
\left(r+1\right)^{\frac{k}{\beta}-1} \Gamma\left(1-\frac{k}{\beta}\right).
\end{eqnarray*}

\noindent The moment generating function $M(z)$ of $T$ for $|z|<1$ is,
\begin{equation*}
M_{t}\left(z\right)=\left\{ \sum^{n}_{k=0} \frac{z^k}{k!} b c^k
\sum^{\infty}_{r=0} \frac{\left(-1\right)^r
  \Gamma\left(b\right)}{\Gamma\left(b-r\right) r!}  
\left(r+1\right)^{\frac{k}{\beta}-1} \Gamma\left(1-\frac{k}{\beta}\right)\right\}.
\end{equation*}

\noindent Hence, for $|z| < 1$, the cumulative generating function of $T$ is
\begin{equation*}%\label{eq_cumulant}
K\left(z\right)=\log \left[ \sum^{n}_{k=0} \sum^{\infty}_{r=0}
 \left[\frac{z^k}{k!} b c^k  \frac{\left(-1\right)^r
 \Gamma\left(b\right)}{\Gamma\left(b-r\right) r!}
 \left(r+1\right)^{\frac{k}{\beta}-1}
 \Gamma\left(1-\frac{k}{\beta}\right)\right] \right]. 
\end{equation*}
We note that it was necessary to use the expansions previously 
presented for the results of this section. 

\subsection{Mean deviations}\label{Meandeviations}

The amount of scattering in a population may be measured by all the 
absolute values of the deviations from the mean or the median. If X 
is a random variable with Kum-IW distribution with mean $\mu=E[X]$ and 
median $M$, then the average deviation from the average and the average 
deviation to the median are defined respectively by,
$$
\delta_{1}(X) = \int_{0}^{\infty} |x-\mu|f(x)dx
\quad\mbox{and}\quad
\delta_{2}(X) = \int_{0}^{\infty} |x-M|f(x)dx.
$$
Using the density of extended Kum-IW and given that,
\begin{equation*}
\int_{\mu}^{\infty} x f(x)dx = bc \sum_{r=0}^{\infty}
\frac{(-1)^{r}\Gamma(b)}{\Gamma(b-r)r!}
\left(\frac{1}{1+r}\right)^{\frac{1}{\beta}-1}  
\gamma\left(\frac{\beta-1}{\beta},\frac{c^{\beta}}{\mu^{\beta}}(1+r)\right),
\end{equation*}
where $\gamma(a,x)$ is the lower incomplete Gamma function, it follows that
\begin{equation*}
\delta_{1}(X) = 2{\mu} F(\mu) - 2\mu + 2bc \sum_{r=0}^{\infty}
\frac{(-1)^{r}\Gamma(b)}{\Gamma(b-r)r!}
\left(\frac{1}{1+r}\right)^{\frac{1}{\beta}-1}  
\gamma\left(\frac{\beta-1}{\beta},\frac{c^{\beta}}{\mu^{\beta}}(1+r)\right)
\end{equation*}
and
\begin{equation*}
\delta_{2}(X) = 2{\mu} F(\mu) - 2\mu + 2bc \sum_{r=0}^{\infty}
\frac{(-1)^{r}\Gamma(b)}{\Gamma(b-r)r!}
\left(\frac{1}{1+r}\right)^{\frac{1}{\beta}-1}  
\gamma\left(\frac{\beta-1}{\beta},\frac{c^{\beta}}{M^{\beta}}(1+r)\right).
\end{equation*}

\subsection{Bonferroni and Lorenz curves}\label{Bonferroni}

Bonferroni and Lorenz curves are widely applied not only in economics 
to study income and poverty, but also in other fields such as reliability, 
demography, insurance and medicine.

\noindent Let then $\mu = E(X)$ e $q = F^{-1}(p;\theta)$, where $F^{-1}(.)$ is the inverse 
function of the cumulative function of a random variable X. Bonferroni 
and Lorenz curves are defined by,
$$
B(p) = \frac{1}{p{\mu}} \int_{0}^{q} x f(x)dx
\quad\mbox{and}\quad
L(p) = \frac{1}{\mu}\int_{0}^{q} x f(x)dx.
$$ 

\noindent Then, using the expanded density,
\begin{equation*}
\int_{0}^{q} x f(x)dx = bc \sum_{r=0}^{\infty}
\frac{(-1)^{r}\Gamma(b)}{\Gamma(b-r)r!}
\left(\frac{1}{1+r}\right)^{\frac{1}{\beta}-1}  
\Gamma\left(\frac{\beta-1}{\beta},\frac{c^{\beta}}{q^{\beta}}(1+r)\right),
\end{equation*}
where $\Gamma(a,x)$ is the upper incomplete gamma function.
Therefore, we have,
\begin{itemize}
\item the Bonferroni curve which is given by: 
\begin{equation*}
B(p) = \frac{bc}{p{\mu}} \sum_{r=0}^{\infty}
\frac{(-1)^{r}\Gamma(b)}{\Gamma(b-r)r!}
\left(\frac{1}{1+r}\right)^{\frac{1}{\beta}-1}  
\Gamma\left(\frac{\beta-1}{\beta},\frac{c^{\beta}}{q^{\beta}}(1+r)\right);
\end{equation*}
\end{itemize}

\begin{itemize}
\item the Lorenz curve which is given by:
\begin{equation*}
L(p) = 
\frac{bc}{{\mu}} \sum_{r=0}^{\infty}
\frac{(-1)^{r}\Gamma(b)}{\Gamma(b-r)r!}
\left(\frac{1}{1+r}\right)^{\frac{1}{\beta}-1}  
\Gamma\left(\frac{\beta-1}{\beta},\frac{c^{\beta}}{q^{\beta}}(1+r)\right).
\end{equation*}
\end{itemize}

\subsection{Order statistics and Shannon entropy}\label{sec_ord}

Let $T_{1:n}\leq T_{2:n}\leq\cdots\leq T_{n:n}$ be the order statistics 
obtained from the Kum-IW distribution. The random variable \textit{$T_{r:n}$}, for 
\textit{$r=1,\ldots,n$}, denotes the $r$-th order statistic in a sample of size $n$. 
The pdf of \textit{$T_{r:n}$} written as,
\begin{eqnarray*}
f_{r:n}(t) = C_{r:n}F(t)^{r-1}\left[1-F(t)\right]^{n-r}f(t),\quad t>0,
\end{eqnarray*}
where $f(t)$ and $F(t)$ are given by equations (\ref{eq_f-Kum-IW}) and
(\ref{eq_Kum-IW}) and $C_{r:n} = n!/[(r-1)! (n-r)!]$. 

The $k$-th moment $\mu^{(k)}_{r:n}$ of the $r$th order statistic is
\begin{eqnarray*}
\mu^{(k)}_{r:n}=E(T^{k}_{r:n})=t^{k} C_{r:n}\int\limits^{\infty}_{0}
\left[F(t)\right]^{r-1}\left[1-F(t)\right]^{n-r}f(t) dt,
\end{eqnarray*}
for $k =1,2,\ldots$, and $1\leq r \leq n$.
Hence,
\begin{equation*}
\mu^{(k)}_{r:n}=C_{r:n} \int^{1}_{0} j^{n-r} \left(1-j\right)^{n-1} dj.
\end{equation*}

\noindent and we obtain an expression for the moment given by,
\begin{equation*}
E(T^{k}_{r:n})=C_{r:n} b c^{\beta} \sum^{\infty}_{j=0} \sum^{\infty}_{i=0} \frac{\left(-1\right)^{i+j} 
\Gamma\left(bn-br+bj+b\right)}{i! j! \Gamma\left(r-j\right) \Gamma\left(bn-br+b-i\right)}
\left(i+1\right)^{\frac{k}{\beta}-1} \Gamma\left(1-\frac{k}{\beta}\right).
\end{equation*}
\vskip .2cm

The Shannon entropy of a random variable $T$ is defined as a measure of the quantity of information. 
A certain message has more quantity of information the greater degree
of uncertainty and is defined mathematically 
by $E\{- \log[f(t)] \}$, where $f(t)$ is the fdp of $T$. In
particular, for a random variable $T$ which follows the Kum-IW distribution we have,
\begin{eqnarray}
&&E\left\{- \log[f(t)] \right\} =
%&=&
-\log \left(\beta b c^{\beta}\right)+
b \left(\beta +1\right) \log(c)\sum\limits_{r=0}^{\infty}
\frac{\left(-1 \right)^r \Gamma\left(b\right)}{\Gamma\left(b-r\right)
  r!}\frac{1}{\left(r+1\right)}- \nonumber\\
&&
\frac{1}{\beta} \sum\limits_{r=0}^{\infty} \frac{\left(-1 \right)^r
  \Gamma\left(b\right)}{\Gamma\left(b-r\right)
  r!}\frac{1}{\left(r+1\right)}\left[\gamma-\log(r+1)\right]+ 
c^{\beta-1} \sum\limits_{r=0}^{\infty} \frac{\left(-1 \right)^r
  \Gamma\left(b\right)}{\Gamma\left(b-r\right)
  r!}\left(r+1\right)^{\frac{1}{\beta}-1}\Gamma\left(1-\frac{1}{\beta}\right)+ \nonumber\\
&&\left(b+1\right) b \sum^{\infty}_{r=0} \sum^{\infty}_{i=0} \frac{\left(-1\right)^{i+r} 
\Gamma\left(b\right) \Gamma\left(b+1\right)}{i! r! \Gamma\left(b-r\right) \Gamma\left(b+1-j\right)} 
\left(j+1\right)^{2}, \quad j > -1
\end{eqnarray}
where $\gamma=-\int^{\infty}_{0}\log(j) \exp\left\{-j\right\} dj$ is the approximate value of the Euler's constant.

\subsection{R\'enyi entropy}\label{RE}

The entropy of a random variable $X$ with density function (\ref{1}) 
measuring the uncertainty of the variation. The R\'enyi entropy is given by,
\begin{equation*}
I_{r}\left(\rho\right)=\frac{1}{1-\rho} \log\left\{\int f\left(x\right)^{\rho} dx\right\}.
\end{equation*}
where $\rho > 0$ and $\rho \neq 1$.

In information theory, R\'enyi entropy generalizes the Shannon entropy.
This form of entropy is important especially in ecology and
statistics, where it can be used 
as an index of diversity. In quantum information, it can be used as a measure 
of entanglement.
If X is a random variable and follows the Kum-IW distribution, then the R\'enyi 
entropy is given by
\begin{equation*}
I_{r}\left(\rho\right)=\frac{1}{1-\rho} \log\left\{ \beta^{\rho-1} b
c^{1-\rho} \sum^{\infty}_{i=0} \frac{\left(-1\right)^{i}  
\Gamma\left(\rho b-\rho+1\right)}{\left(i !\right) \Gamma\left(\rho b-\rho+1-i\right)} 
\left(r+i\right)^{\frac{1}{\beta}-\rho\left(1+\frac{1}{\beta}\right)}
\Gamma\left(\frac{\rho}{\beta}+\rho-\frac{1}{\beta}\right)\right\}. 
\end{equation*}

\section{Inference for Censored Data}\label{sec:estimation}

\subsection{Maximum likelihood estimation}\label{sec_mle}

Let $T_i$ be a random variable with Kum-IW distribution with parameter vector 
$\bm{\theta} = (b,c,\beta)$. The data in survival analysis and reliability studies are
generally censored. A very simple random censoring mechanism that is
often realistic is one in which each individual $i$ is assumed to have
a lifetime $T_i$ and a censoring time $C_i$, where $T_i$ and $C_i$ are
independent random variables. Suppose the data set consists of $n$
independent observations $t_i=\mbox{min}(T_i,C_i)$ for
$i=1,\cdots,n$. The distribution of $C_i$ does not depend on any of
the unknown parameters of $T_i$. Parametric inference for such data
are usually based on likelihood methods and their asymptotic
theory. The censored log-likelihood $l(\tetn)$ for the model
parameters is 

\begin{equation*}
\ell(\tetn)=r \log\left(\beta b c^\beta\right)-c^\beta \sum_{i \in F}
\left( \frac{1}{t_{i}} \right)^\beta-\left(\beta+1\right) \sum_{i \in
  F} \log(t_{i}) 
+\left(b-1\right)\sum_{i \in F} \log \left[ 1-\exp \left[ -\left( \frac{c}{t_{i}} \right)^\beta \right] \right]+
\end{equation*}
\begin{equation}
b \sum_{i \in C} \log \left[ 1-\exp \left[ -\left( \frac{c}{t_{i}} \right)^\beta \right] \right],
\end{equation}
where $F=\left[1, r\right]$ and $C=\left[r+1, n\right]$; still, $C$
represents the censored data and  $F$ represents the failure data. 

The maximum likelihood estimate (MLE) $\widehat\tetn$ of $\tetn$ is
obtained by solving the nonlinear likelihood equations
$U_{b}(\tetn)=\frac{\partial\ell\left(\tetn\right)}{\partial b}=0$,
$U_{c}(\tetn)=\frac{\partial\ell\left(\tetn\right)}{\partial c}=0$ and
$U_{\beta}(\tetn)=\frac{\partial\ell\left(\tetn\right)}{\partial
  \beta}=0$. These equations cannot be solved analytically and
statistical software can be used to solve the equations numerically.  

For interval estimation of $b$, $c$ and $\beta$, and tests of
hypotheses on these parameters, we must obtain the $3 \times 3$ observed information
matrix $J(\tetn)$ which is given by,
\begin{displaymath}
\mathbf{J\left(\theta\right)}=
\left( \begin{array}{ccc}
J_{b b}\left(\theta\right) & J_{b c}\left(\theta\right) & J_{b \beta}\left(\theta\right)\\  
J_{c b}\left(\theta\right) & J_{c c}\left(\theta\right) & J_{c \beta}\left(\theta\right)\\
J_{\beta b}\left(\theta\right) & J_{\beta c}\left(\theta\right) & J_{\beta \beta}\left(\theta\right)\\
\end{array}\right),
\end{displaymath}

Under conditions met for parameters obeying the parametric space and
not considering the limits of the same, the asymptotic distribution
of $$\sqrt n
(\widehat\tetn-\tetn)\,\,\,\,\mathrm{is}\,\,\,\,N_3(0,I(\tetn)^{-1}),$$
where $I(\tetn)$ is the expected information matrix. This asymptotic
behavior is valid if $I(\tetn)$ is replaced by $J(\widehat\tetn)$, the
observed information matrix evaluated at $\widehat\tetn$. The
asymptotic multivariate normal distribution
$N_3(0,J(\widehat\tetn)^{-1})$ can be used to construct approximate
confidence intervals and confidence regions for the individual
parameters and for the hazard and survival functions. The asymptotic
normality is also useful for testing goodness of 
fit of the three parameters the Kum-IW distribution and for comparing
this distribution with some of its special submodels using one of the
two well-known asymptotically equivalent test statistics - namely, the
likelihood ratio (LR) statistic and the Wald and Rao score
statistics. 

\subsection{Bayesian approach}\label{prior}

Following the Bayesian paradigm, we need to complete the model
specification by specifying a prior
distribution for the parameters. By Bayes Theorem, the
posterior distribution is then proportional to the product of the
likelihood function by the prior density.

Subjetivism is the predominant philosophical foundation in
Bayesian inference, although in practice noninformative prior densities
(built on some formal rule) are frequently used 
(\citealt{Kass1996}). Since the parameters in the Kum-IW distribution
are all positive quantities and due to the flexibility generated by
the two-parameter Gamma distribution this is adopted as prior
distribution. So, $b\sim Gamma(m,w)$, $c\sim Gamma(a,s)$ and 
$\beta\sim Gamma(x,l)$. 

Assuming independence among the prior
densities, the posterior density is expressed by,
\begin{eqnarray}\label{priorversuslikelihood}
h(\bm{\theta}|t) &\propto&
b^{m+n-1} c^{a+n\beta-1} \beta^{x+n-1}
\exp\left[-wb-sc-l \beta-c^{\beta} \sum^{n}_{i=1} t^{-\beta}_{i}\right]\nonumber\\
&&
\left\{\prod^{n}_{i=1} t^{-\beta-1}_{i} \left\{1-
\exp\left[-\left(\frac{c}{t_{i}}\right)^{\beta}\right]\right\}^{b-1}
\right\}.  
\end{eqnarray}

\noindent This joint density has no known analytical form but we can
provide an
approximate solution based on the complete conditional
distributions of $b$, $c$ and $\beta$. These are given by the following expressions,

\begin{equation*}
h(b \mid  c, \beta, t) \propto b^{m+n-1} 
\exp\left[-wb-c^{\beta} \sum^{n}_{i=1} t^{-\beta}_{i}\right]
\left\{\prod^{n}_{i=1} t^{-\beta-1}_{i} \left\{1-
\exp\left[-\left(\frac{c}{t_{i}}\right)^{\beta}\right]\right\}^{b-1}
\right\},  
\end{equation*} 

\begin{equation*}
h(c \mid b, \beta, t) \propto  c^{a+n\beta-1} 
\exp\left[-sc-c^{\beta} \sum^{n}_{i=1} t^{-\beta}_{i}\right]
\left\{\prod^{n}_{i=1} t^{-\beta-1}_{i} \left\{1-
\exp\left[-\left(\frac{c}{t_{i}}\right)^{\beta}\right]\right\}^{b-1}
\right\},
\end{equation*}

\begin{equation*}
h(\beta \mid b, c, t) \propto \beta^{x+n-1}
\exp\left[-l \beta-c^{\beta} \sum^{n}_{i=1} t^{-\beta}_{i}\right]
\left\{\prod^{n}_{i=1} t^{-\beta-1}_{i} \left\{1-
\exp\left[-\left(\frac{c}{t_{i}}\right)^{\beta}\right]\right\}^{b-1}
\right\}.
\end{equation*}

\section{Application}\label{Application}

In this section, we present estimation results for
parameters of the Kum-IW
distribution under a Bayesian approach. The 
commercial production of cattle meat
in Brazil, which usually comes from the cattle of the
Nelore race, search to optimize the process trying to obtain a time
for the cattle to reach the specific weight in the period of the
birth until it weans. For a data set with 69 bulls of the Nelore race,
was observed the time (in days) until the animals reach the weight
of 160kg relative to the period from birth until it weans. We 
compared the Kaplan-Meyer and Bayesian survival functions through 
two graphic methods.

Using the expression of $\ell(\theta)$ which a routine was escribed
in the software Winbugs see \cite{Spiegelhalter2007} to
esteem the values of the vector of parameters $\theta=(b,
c, \beta)^{T}$ of the Kum-IW distribution
for the data of the cattle of the Nelore race.

\begin{table}[!htb]\label{tete}
\centering {\caption{Results of the Bayesian approach to
Kum-IW distribution}\vspace*{0.5cm}\label{EMV1}
\begin{tabular}{cccccccc}
\hline Parameter & \qquad Mean  & \qquad SD     & \qquad 2.5\%  & \qquad Median & \qquad 97.5\%\\
\hline
b         	 & \qquad 6.656 & \qquad 11.18  & \qquad 0.9829 & \qquad 3.82 & \qquad 29.76\\
c         	 & \qquad 177.5 & \qquad 19.18  & \qquad 154.9  & \qquad 172.8& \qquad 227.5\\
$\beta$          & \qquad 8.231 & \qquad 2.917  & \qquad 3.84   & \qquad 7.812& \qquad 15.01\\
\hline
\end{tabular}}
\end{table}

%\begin{center}
%Table \ref{tete} about here.
%\end{center}

\begin{figure}[h]\centering
\subfloat[\label{KM1}]{\includegraphics[width=6.5cm,height=6.5cm]{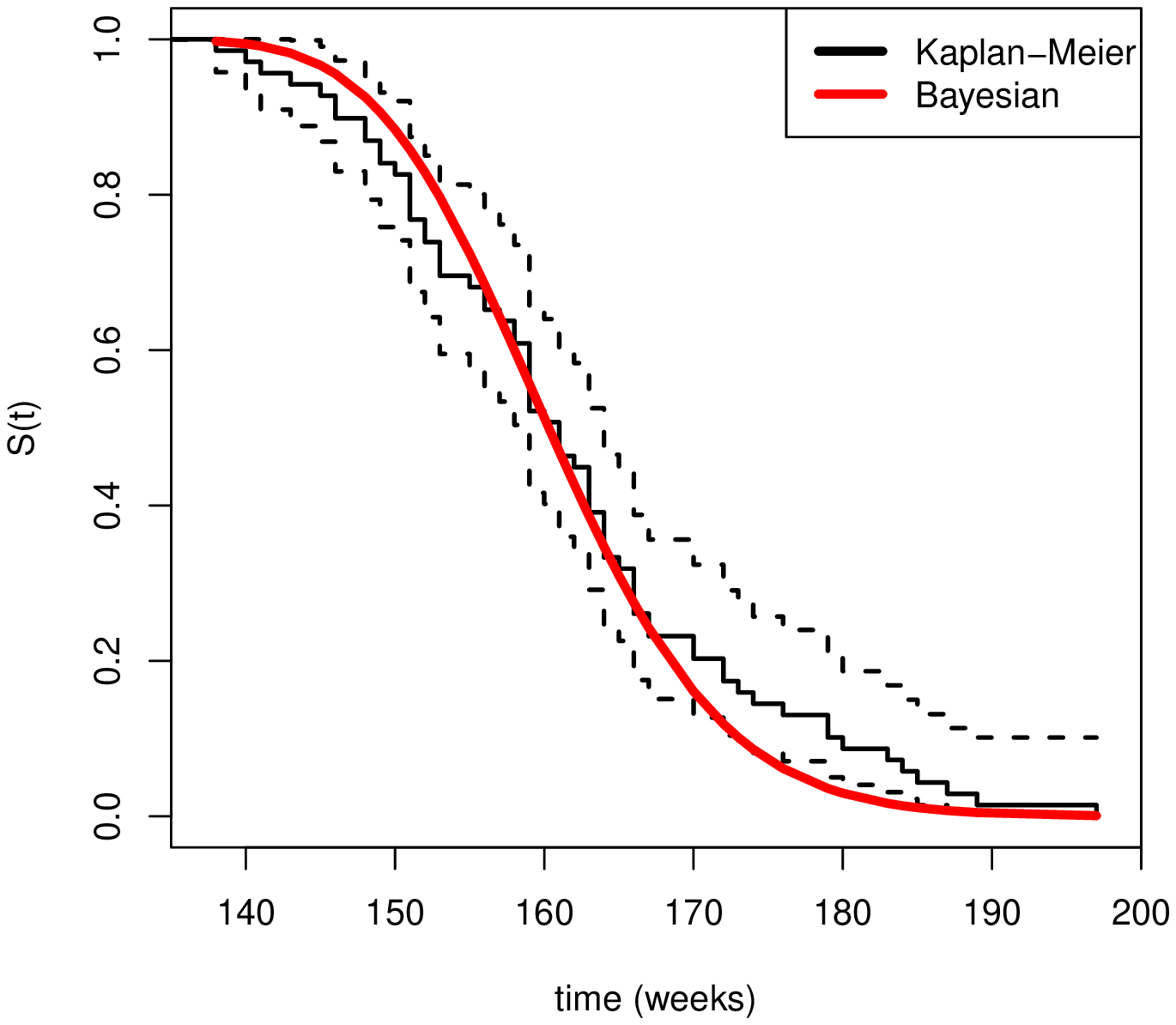}}
\subfloat[\label{KM2}]{\includegraphics[width=6.5cm,height=6.5cm]{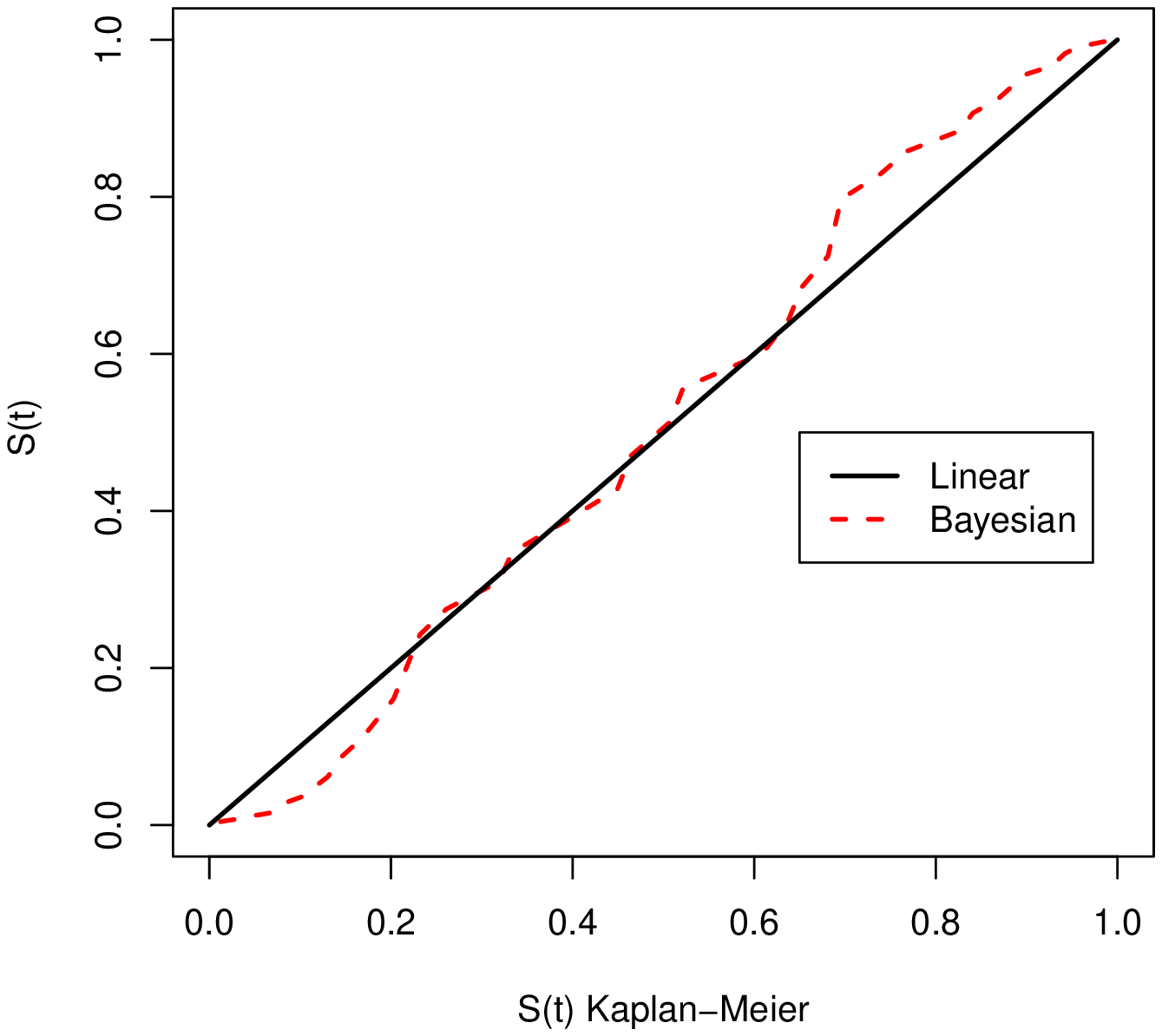}}
\caption{\label{KMT} (a) Comparison among survivals generated by the method Bayesian and
Kaplan-Meier described in the graph S(t) versus time. (b) Comparison among survivals
generated by the method Bayesian described in the graph S(t) versus Kaplan-
Meier.}
\end{figure}

One of
the methods that dispose for us to test if our model is well
adjusted to the data consists in the comparison of the function of
survival of the parametric model proposed with the estimator of
Kaplan-Meier. Another method consists of sketching the survival function 
of the model parametric versus the estimate of Kaplan-Meier
for the survival function, if this curves is close of the straight
line $y=x $ we will have a good adjustment, figure\ref{KMT}.

\setkeys{Gin}{width=6in,height=5in}
\begin{figure}[h]\centering
\includegraphics{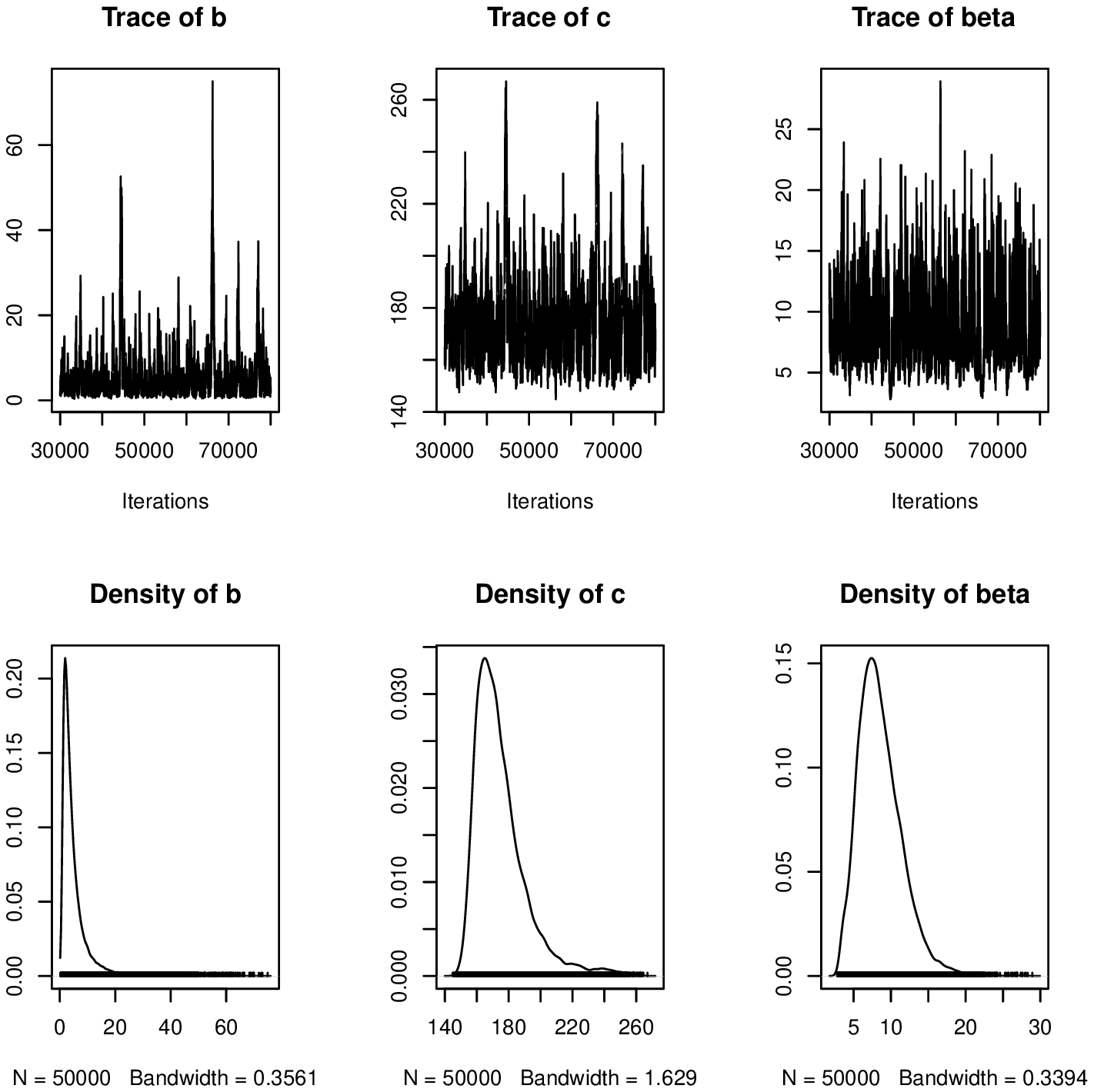}
\caption{Trace plots of simulated parameter values of the Kumaraswamy inverse Weibull distribution.}
\label{fig4}
\end{figure}

%\begin{center}
%Figure \ref{fig4} about here.  
%\end{center}

%\begin{center}
%Figure \ref{KMT} about here.  
%\end{center}

\section{Conclusions}\label{sec_final}

We worked a three parameter lifetime distribution called the Kumaraswamy Inverse 
Weibull (Kum-IW) distribution which extends Inverse Weibull distribution proposed 
and widely used in the lifetime literature. The model is much more flexible than 
the inverse Weibull. The Kum-IW distribution could have increasing, decreasing 
and unimodal hazard rates. We provide a mathematical overview of this distribution 
including the densities of the order statistics, R\'enyi entropy, Shannon entropy, 
Bonferroni and Lorenz curves and Mean deviations. Also, we derive an  
explicit algebraic formula for the $r$-th moment, expressions for the order 
statistics, and the maximum likelihood estimation for the censored data. The 
performance of the model was analized using real data sets where the Kum-IW 
distribution performed verywell and the estimation was given by Bayes method.

\bibliographystyle{natbib}

\end{document}